\documentclass[letter]{jpsj3}

\usepackage{amsmath}

\title{Possible Long-Range Order with Singlet Ground State in CeRu$_2$Al$_{10}$}

\author{
Hiroshi \textsc{Tanida}\thanks{E-mail address: tany@hiroshima-u.ac.jp}, 
Daiki \textsc{Tanaka}, Masafumi \textsc{Sera}, Chikako \textsc{Moriyoshi}$^1$, Yoshihiro \textsc{Kuroiwa}$^1$,\\ Tomoaki \textsc{Takesaka}$^2$, Takashi \textsc{Nishioka}$^2$, Harukazu \textsc{Kato}$^2$ and Masahiro \textsc{Matsumura}$^2$
}

\inst{
Department of Quantum Matter, ADSM, Hiroshima University, Higashi-Hiroshima, Hiroshima, Japan 739-8530

$^1$Department of Physics, Hiroshima University, Higashi-Hiroshima, Hiroshima, Japan 739-8530

$^2$Graduate School of Integrated Arts and Science, Kochi University, Kochi, Japan 780-8520
}

\abst{
We investigate the thermal and transport properties of Ce$_x$La$_{1-x}$Ru$_2$Al$_{10}$ to clarify the origin of the recently discovered mysterious phase below $T_0$=27 K in CeRu$_{2}$Al$_{10}$ where a large magnetic entropy is released, however, the existence of an internal magnetic field is ruled out by $^{27}$Al-NQR measurement.
We find that $T_0$ decreases with decreasing $x$ and disappears at $x\sim$0.45. 
$T_0$ of CeRu$_2$Al$_{10}$ is suppressed down to 26 K under $H$=14.5 T along the $a$-axis. 
These results clearly indicate that the transition has a magnetic origin and is ascribed to the interaction between Ce ions.
Considering the results of specific heat, magnetic susceptibility, thermal expansion, and electrical resistivity and also $^{27}$Al-NQR, we propose that the transition originates from the singlet pair formation between Ce ions. 
Although its properties in a Ce dilute region is basically understood by the impurity Kondo effect, CeRu$_2$Al$_{10}$ shows a Kondo-semiconductor-like behavior.
The phase transition at $T_0$ may be characterized as a new type of phase transition that appears during the crossover from the dilute Kondo to the Kondo semiconductor.

}

\kword{CeRu$_2$Al$_{10}$, spin gap, singlet ground state, Kondo semiconductor}

\begin{document}
\maketitle

The ternary rare-earth compound CeRu$_2$Al$_{10}$ with an orthorhombic structure \cite{Verena,Tursina} shows a mysterious transition at $T_0$=27 K. \cite{Strydom, Takesaka, Nishioka, Matsumura}
In the early stage of the study, the origin of the transition was considered to be the antiferromagnetic (AFM) ordering from a cusp of magnetic susceptibility and a large peak of specific heat at $T_0$. \cite{Strydom}
However, $T_0$ is too high for the series of $R$Ru$_2$Al$_{10}$ ($R$=rare-earth element). 
The highest magnetic ordering temperature is 16.5 K in GdRu$_2$Al$_{10}$. \cite{Takesaka, Nishioka}
Assuming the de Gennes law, the transition temperature of CeRu$_2$Al$_{10}$ is evaluated to be much lower than 27 K. 
The distance between neighboring Ce ions is as large as ${\sim}$5.2 ${\rm {\AA}}$. \cite{Tursina}
Thus, we proposed that $T_0$ is not the AFM ordering temperature. \cite{Nishioka, Takesaka}

In our previous papers,\cite{Nishioka, Takesaka} we reported the transport and magnetic properties of CeRu$_2$Al$_{10}$ single crystals, and suggested the CDW transition as a possible candidate for the transition in the following reasons. \cite{Nishioka}
The specific heat shows a sharp peak at $T_0$ and the temperature dependence below $T_0$ is expressed as $exp(-{\Delta}/T)$, where ${\Delta}$ is the excitation gap. 
Electrical resistivity was explained by considering the fact that the partial gap on the Fermi surface is formed below $T_0$. 
The $exp(-{\Delta}/T)$ like behavior below $T_0$ was also observed for electrical resistivity and magnetic susceptibility.
We also reported the results of $^{27}$Al-NQR where sharp peaks from five nonequivalent Al sites were observed, and, from the very simple splitting of the peaks in the NQR spectrum below $T_0$, the usual magnetic ordering was ruled out. \cite{Matsumura} 
Then, the nonmagnetic structural transition was considered as another candidate of the transition.\cite{Matsumura}
The unusual property showing the decrease in the magnetic susceptibility along all the crystal axes below $T_0$\cite{Nishioka} remains to be explained.

Thus, no definite conclusion has yet been obtained for the long-range order (LRO) below $T_0$. 
The fact that the magnetic entropy at $T_0$ is large suggests the possibility that the crystalline-electric-field (CEF) ground doublet is associated with it. 
No LRO exists in LaRu$_2$Al$_{10}$. 
This suggests that the interaction between Ce ions is associated with the LRO. 
If this is the case, the suppression of $T_0$ by La doping and the magnetic field effect are expected. 
In this letter, we performed the study of Ce$_x$La$_{1-x}$Ru$_2$Al$_{10}$ and the magnetic field effect in order to clarify the nature of the LRO.

Single crystals were prepared by the Al self-flux method.  
The specific heat, $C$ was measured using PPMS. 
The magnetic susceptibility, ${\chi}$ was measured using MPMS.
The thermal expansion, ${\Delta}l/l$ was measured by the three-terminal capacitance method.
The electrical resistivity, ${\rho}$ was measured by the usual four-probe ac-method.
The powder X-ray diffraction of CeRu$_2$Al$_{10}$ was examined using BL02B2 in SPring-8. 
The lattice constants $a$, $b$, and $c$ of CeRu$_2$Al$_{10}$ at room temperature were confirmed to be 9.12679(4), 10.28036(4), and 9.18798(4) ${\rm {\AA}}$ by Rietveld analysis, respectively. 
These lattice constants suggest the trivalency of Ce ions.

Figure 1 shows the temperature, $T$ dependence of the magnetic part of $C$ of Ce$_x$La$_{1-x}$Ru$_2$Al$_{10}$ ($x$=1, 0.9, 0.7, 0.5, 0.3, 0.1) in the form $C_{4f}/T$.
Here, the lattice contribution is subtracted using that of LaRu$_2$Al$_{10}$. 
A sharp peak of $C_{4f}/T$ at $T_0$ is rapidly smeared out only by a weak La doping and, $T_0$ decreases down to $\sim$6 K at $x$=0.5.
However, the magnitude of the released magnetic entropy at $T_0$, which is ${\sim}$0.65 $R$ln2 for $x$=1, is almost independent of $x$ down to $x$=0.5.
The decrease in $T_0$ with decreasing $x$ clearly indicates that the transition at $T_0$ should not be ascribed to the nonmagnetic origin such as CDW and structural transition but to the magnetic origin from the interaction between Ce ions.
We note that the critical concentration of $x_c\sim$0.45 for the phase transition is not unusual for rare-earth compounds in general.
Although there is as yet no detailed information on the CEF level scheme at present, the large anisotropy in ${\chi}$ suggests that the overall CEF splitting is expected to be more than ${\sim}$300 K and that the CEF ground state is considered to be a well isolated doublet.\cite{Nishioka}
The magnetic entropy at $T_0$ is expected to originate from this ground doublet. 
For $x$=0.3 and 0.1, no transition is observed down to 0.4 K. 
On the other hand, $C_{4f}/T$ markedly increases with decreasing temperature. 
Together with the results of ${\rho}$ which will be shown later, the low-temperature increase in the $C_{4f}/T$ of these dilute compounds is ascribed to the Kondo effect.

\begin{figure}[tb]
\begin{center}
\includegraphics[width=0.58\linewidth]{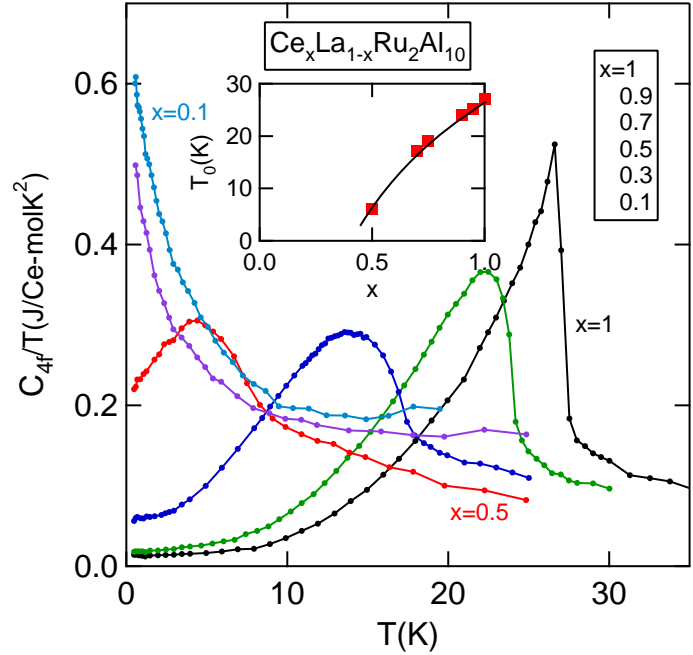}
\end{center}
\caption{
(Color online) Temperature dependence of the specific heat of Ce$_x$La$_{1-x}$Ru$_2$Al$_{10}$ in the form of $C_{4f}/T$ at $H$=0. 
The inset shows the $x$ dependence of the transition temperature $T_0$.
}
\label{f1}
\end{figure}

\begin{figure}[tb]
\begin{center}
\includegraphics[width=0.58\linewidth]{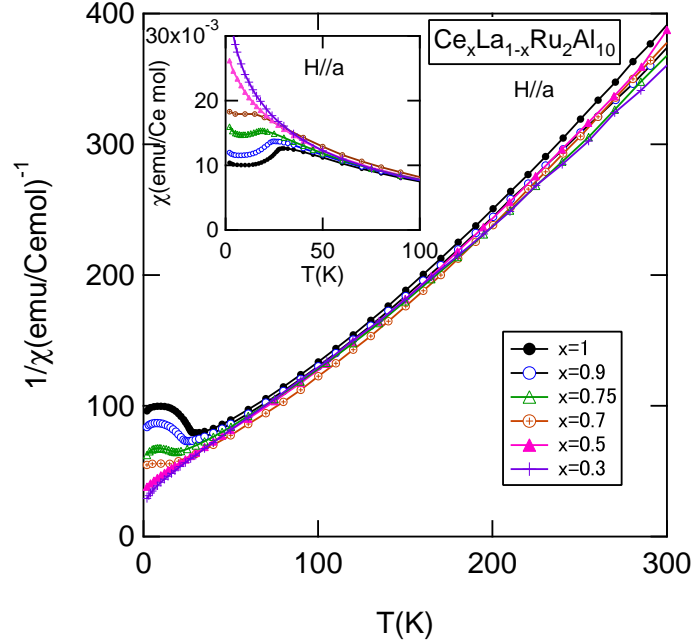}
\end{center}
\caption{
(Color online) Temperature dependence of the inverse magnetic susceptibility of Ce$_x$La$_{1-x}$Ru$_2$Al$_{10}$ measured at $H$=1 T along the $a$-axis.\cite{Nishioka2}  The inset shows that of the magnetic susceptibility at low temperatures.
}
\label{f1}
\end{figure}

Figure 2 shows the $T$ dependence of the inverse magnetic susceptibility ${\chi}^{-1}$ of Ce$_x$La$_{1-x}$Ru$_2$Al$_{10}$ measured at $H$=1 T along the $a$-axis.\cite{Nishioka2}
The inset shows that of ${\chi}_a$ below 100 K. 
As seen in the inset of Fig. 2, $T_{0}$ decreases with decreasing $x$, consistent with the results of $C$.
The decrease in ${\chi}_a$ below $T_0$ is strongly suppressed for $x<$0.8. 
As for $x$=0.7, the cusp-like peak is no longer observed at $T_0$. 
The anomaly is found as only a kink. 
It is noted that the low temperature upturn of $\chi_a$ is small even for $x$=0.7. 
This suggests that there are a small number of free magnetic ions and that most of the Ce ions take part in the LRO below $T_0$.
For $x$=0.5, although the transition is recognized in $C$, it is not recognized in ${\chi}_a$ and only a Curie-like upturn is seen at low temperatures.
The $T$ dependence of $\chi_a$ above $T$${\sim}$100 K is weak at $x$. 
This means that the character of Ce ion does not vary with $x$ at high temperatures. 
An upward deviation from the $T$ linear dependence seen in high-temperature region appears at $T$${\sim}$130 K in ${\chi}_a^{-1}$.
At present, although we do not know its origin, the CEF effect is one of the possible origins.
Apart from the CEF effect, below $T$${\sim}$100 K, a weak $x$ dependence is seen, namely, the suppression of ${\chi}_a$ with increasing $x$. 
This suggests the development of the AFM correlation below $T$$\sim$ 100 K in compounds with large $x$ values. 
The magnetization $M$ along the $a$-axis of $x$=1 at $T$=1.4 K increases almost linearly to $H$ and is 0.3 ${\mu}_{\rm B}/$Ce at $H$=14.5 T.
This suggests that the upper critical field to the paramagnetic region is much higher than $H$=14.5 T, if it exists.

\begin{figure}[tb]
\begin{center}
\includegraphics[width=0.56\linewidth]{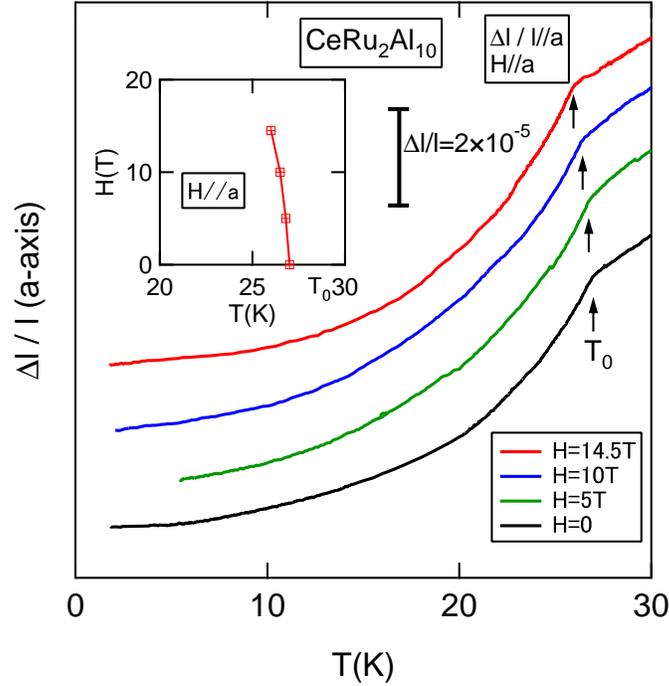}
\end{center}
\caption{
(Color online) Temperature dependence of the thermal expansion along the $a$-axis in longitudinal magnetic fields.
}
\label{f1}
\end{figure}

Figure 3 shows the $T$ dependence of the thermal expansion ${\Delta}l/l$ of CeRu$_2$Al$_{10}$ along the $a$-axis under the longitudinal magnetic fields. 
All the results in magnetic fields show similar trends. 
Namely, with decreasing temperature, ${\Delta}l/l$ shows a monotonic shrinkage with decreasing temperature towards $T_0$ and a steep shrinkage below $T_0$, its shrinkage becomes smaller with further decrease in temperature.  
It is seen that $T_0$ decreases from 27 to 26 K with increasing magnetic field up to 14.5 T as shown in the inset of Fig. 3. 
The decrease in $T_0$ with increasing magnetic field also indicates that the LRO is formed by the interaction between Ce ions, and that the critical field from the LRO to the paramagnetic state should exist at a high field. 
The ${\Delta}l/l$'s along the $b$- and $c$-axes also show similar temperature dependences with the same order of magnitude of shrinkage, but $T_0$ is independent of magnetic field strength, which is expected from the small $M$ along these two directions.
These results indicate that crystal volume decreases below $T_0$.
Namely, the LRO below $T_0$ has an energy gain by reducing crystal volume. 
This is consistent with the enhancement in $T_0$ with pressure application. \cite{Nishioka}

\begin{figure}[tb]
\begin{center}
\includegraphics[width=0.548\linewidth]{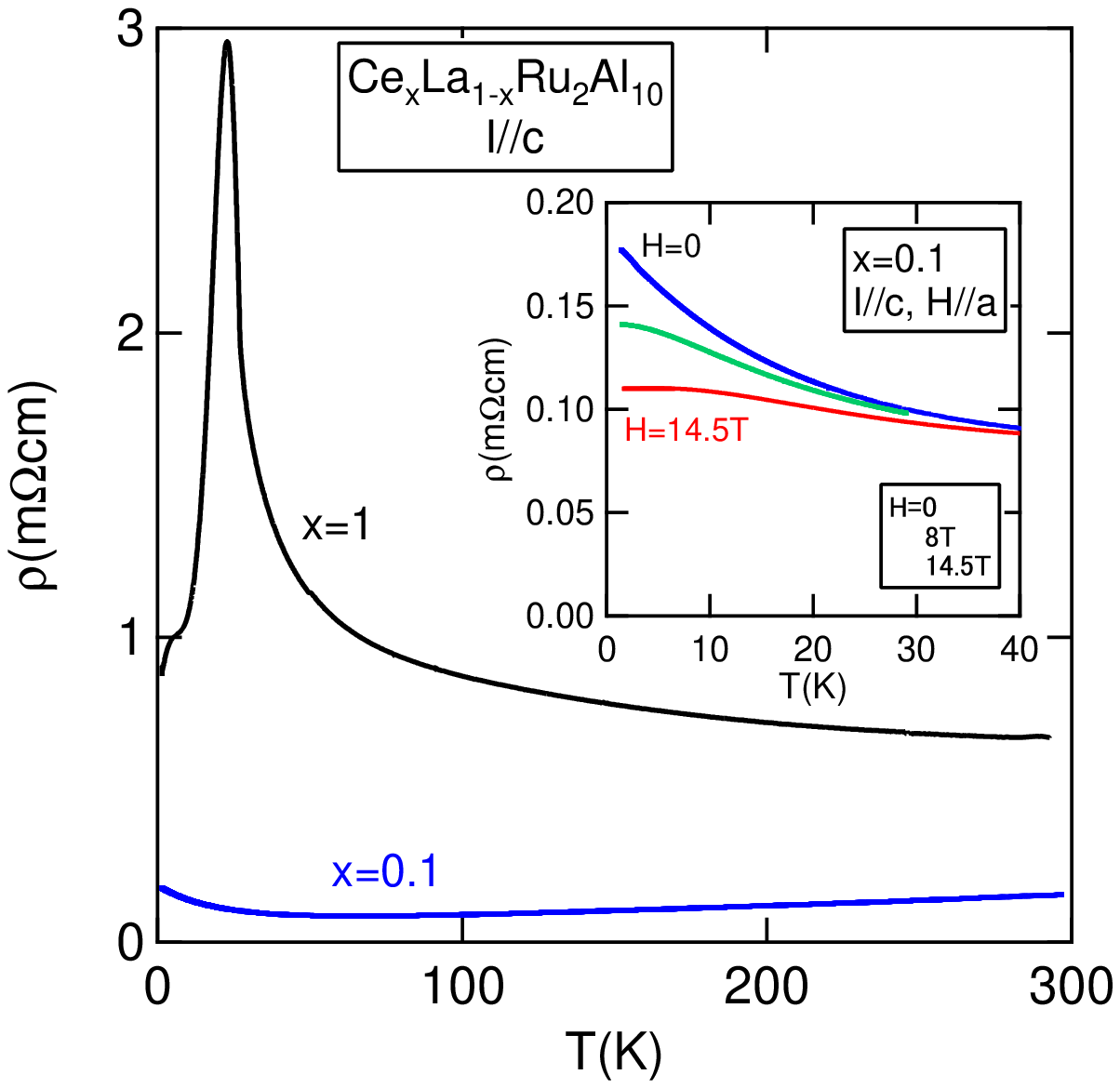}
\end{center}
\caption{
(Color online) Temperature dependence of the electrical resistivity of Ce$_x$La$_{1-x}$Ru$_2$Al$_{10}$($x$=1 and 0.1).  $I\parallel c$ and $H\parallel a$.
}
\label{f1}
\end{figure}

Figure 4 shows the $T$ dependence of $\rho$ of Ce$_x$La$_{1-x}$Ru$_2$Al$_{10}$ ($x$=1, 0.1) at $H$=0 T.
Here, $I\parallel c$. 
For $x$=1, $\rho$ increases with decreasing temperature from room temperature and the rate of increase is enhanced largely below $\sim$50 K.
After showing a sudden increase just below $T_0$, $\rho$ exhibits a pronounced maximum at $T\sim$23 K and then a sharp decrease with dereasing temperature.
A small shoulder is seen at $T\sim$4 K. 
These behaviors are the same as those reported in our papers. \cite{Nishioka, Takesaka}
The absolute value of $\rho$ at low temperature is large and thus the compound should be categorized as a Kondo semiconductor.
A sudden increase in $\rho$ below $T_0$ may be ascribed to the partial gap opening of the Fermi surface. \cite{Nishioka}
A small shoulder at approximately around 4 K suggests the existence of some kind of structure in a low-energy excitation.
For $x$=0.1, $\rho$ shows a metallic behavior at high temperatures but, after showing a minimum at $T\sim$ 50 K, it shows an increase with decreasing temperature.
These behaviors are typical of a single impurity Kondo system.
This is also supported by the large negative magnetoresistance at low temperatures shown in the inset of Fig. 4. 
The result of $C$ for $x$=0.1 showing an increase in $C_{4f}/T$ with decreasing temperature also supports the existence of the Kondo effect.
These indicate that Ce ion is in principle trivalent for $x$=0.1.

\begin{figure}[tb]
\begin{center}
\includegraphics[width=0.548\linewidth]{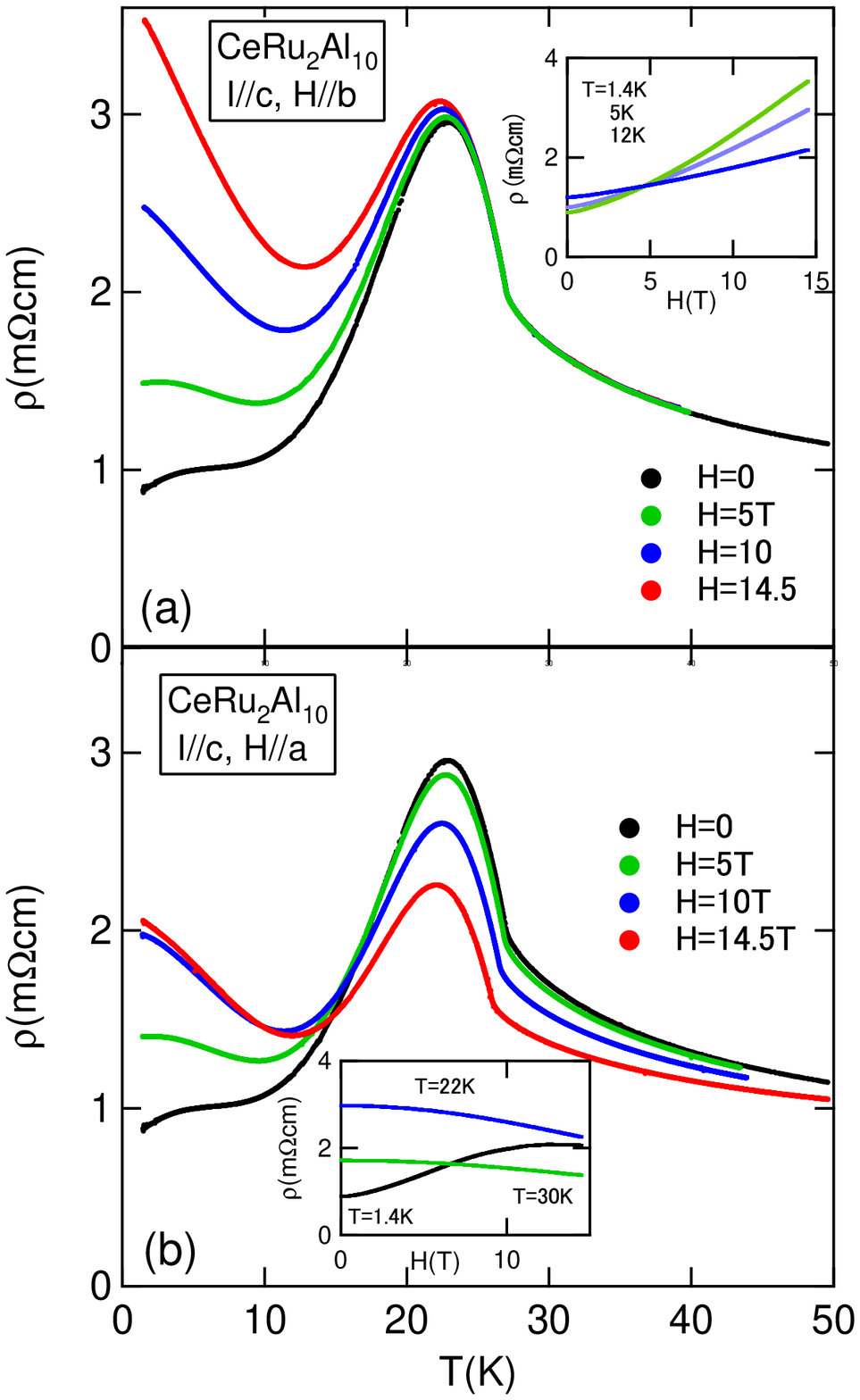}
\end{center}
\caption{
(Color online) Temperature dependence of the electrical resistivity of CeRu$_2$Al$_{10}$ under the magnetic field along (a) $H\parallel b$ and (b) $H\parallel a$.  The insets show the magnetoresistance.
}
\label{f1}
\end{figure}

Figures 5(a) and 5(b) show the $T$ dependences of $\rho$ under the transverse magnetic fields along the $b$- and $a$-axes, respectively.
Here, $I\parallel c$.
The insets show the magnetoresistance. 
For $H\parallel b$, $\rho$ is slightly affected by the magnetic field above $T\sim$ 25 K and $T_0$ is also slightly affected up to 14.5 T.
This robustness of $\rho$ and $T_0$ to the magnetic field at high temperatures should be associated with the small magnetization along the $b$-axis.
Below $T\sim$ 23 K, a positive magnetoresistance is observed, and the lower the temperature, the larger the magnitude of positive magnetoresistance.
As shown in the inset in Fig. 5(a), $\rho$ seems to show a continuous increase up to much higher magnetic fields.
This may be attributed to the cyclotron motion of conduction electron under a transversal magnetic field in a low-carrier system, which was suggested by the result of the Hall effect.\cite{Strydom}
The results for $H\parallel a$ are rather different from those for $H\parallel b$.
This should originate from the large difference in the magnitude of $M$.
Namely, $M$ along the $a$-axis is much larger than that along the $b$-axis.
For $H\parallel a$, above $T\sim$ 15 K, a negative magnetoresistance is observed, which is the largest at $\sim$22 K. 
Even at temperatures as high as 50 K, the observed negative magnetoresistance is still large. 
$T_0$ decreases from 27 K at $H$=0 T to 26 K at $H$=14.5 T, which is consistent with the result of ${\Delta}l/l$. 
This strong magnetic field dependence of $\rho$ originates from the large $M$ along the $a$-axis.
A negative magnetoresistance above $T_0$ should be attributed to the suppression of the scattering of conduction electrons by a localized magnetic moment.
The negative magnetoresistance, which is observed down to $T\sim$ 15 K, is also of the same origin as that mentioned above. 
Below $T\sim$ 15 K, a positive magnetoresistance is observed.
As shown in the inset, at $T$=1.4 K, after showing a broad maximum at $H\sim$ 12 T, ${\rho}$ shows a negative magnetoresistance at higher fields. 
The positive magnetoresistance at low magnetic fields originates from the transverse magnetic field effect for $H\parallel b$, and the origin of the negative one above $\sim$13 T may be the same as that at high temperatures.

In the present study, we found the following important information on the LRO below $T_0$.
From the study of Ce$_x$La$_{1-x}$Ru$_2$Al$_{10}$, it is revealed that $T_0$ decreases gradually and disappears at $x_c$${\sim}$0.45. 
This critical concentration $x_c$ is comparable to that often observed in rare-earth compounds. 
It is also revealed that $T_0$ is suppressed by a magnetic field along the $a$-axis. 
These two results indicate that the origin of the LRO is magnetic.
Namely, the LRO is caused by the interaction between Ce ions. 
The existence of the Kondo effect for $x$=0.1, the large negative magnetoresistance for $H\parallel a$ in a paramagnetic region for $x$=1, and the weak $x$ dependence of $\chi_a-T$ at high temperatures indicate that the Ce ion is in principle trivalent in this system.

Here, we discuss the origin of the transition at $T_0$. 
A magnetic entropy as large as $\sim$0.65$R \ln 2$ is released at $T_0$.
This strongly suggests that the magnetic entropy originates from the CEF ground doublet and that the LRO is caused by some type of magnetic interaction.
As was shown in our previous paper,\cite{Matsumura} in the LRO, there exists no internal magnetic field, which indicates that the ground state is nonmagnetic.
Although the possibility of multipole ordering could be considered, it is ruled out because the CEF ground state is expected to be a well isolated doublet. 
Another important result of our previous study is that $\chi$ along all the crystal axes of $a$, $b$, and $c$ shows a decrease below $T_0$.\cite{Nishioka}
It was also proposed that the unit cell is doubled below $T_0$. \cite{Matsumura}
From these and the present results, we propose that the ground state below $T_0$ is a nonmagnetic singlet such as a spin-Peierls transition,\cite{Hase, Hiroi} although many experimental results remain to be explained from this standpoint, as will be pointed out later.
The excited magnetic state is considered to be situated $\sim$100 K above the ground state, as was estimated in our previous paper.\cite{Nishioka}
From this standpoint, the volume shrinkage below $T_0$ could be naturally explained because in order to construct a singlet ground state, the distance between Ce ions should be decreased. 
The large enhancement in the thermal conductivity ${\kappa}$ below $T_0$ \cite{Strydom} also could be explained as follows. 
The absolute value of ${\rho}$ is large and so ${\kappa}$ is dominated by phonons. 
Below $T_0$, by forming a singlet pair between Ce ions, the vibration of Ce ions is suppressed. 
As a result, the mean free path of phonons is enhanced below $T_0$. 
In order to construct a singlet ground state, in any case, a singlet pair between Ce ions should be formed.
$T_0$ is also observed in the La-doped system down to $x_c\sim$0.45.
This suggests that a singlet pair should be formed not only between nearest neighbors but also between Ce ions whose distance is expanded to the second- or third-nearest neighbors.
The sharp peak of $C$ is rapidly smeared out by La doping but the magnetic entropy below $T_0$ is not markedly changed.
These results suggest that even in La-doped systems, the singlet pair is formed but the magnitude of the spin gap is distributed owing to La doping.

Here, we note the problems to be explained from the present standpoint that the singlet ground state is constructed below $T_0$.
In many insulating 3$d$ compounds, the singlet ground state is often observed and has been extensively studied.
In this case, the low dimensionality of the array of 3$d$ ion plays an essential role. 
However, in the present system, at present, we do not know whether or not the low dimensionality exists in the crystal.
Furthermore, considering the appearance of the same LRO in La-doped systems, it is difficult to conjecture what type of pair is formed.
In the present system, the compound is not insulating but has conduction electrons.
It is expected that conduction electrons will contribute to singlet pair formation.
Another large difference between the 3$d$ and 4$f$ systems is the nonexistence or existence of orbital degrees of freedom.
In 3$d$ systems, as a result of the orbital angular momentum, $L$ is quenched in many cases, and thus the ideal isotropic singlet ground state can be formed.
However, in 4$f$ systems, $L$ exists in a total angular momentum through the spin-orbit coupling.
Then, when the singlet ground state is formed, the existence of $L$ should be contained.
YbAl$_3$C$_3$ is such an example. \cite{Ochiai}
As another example, Yb$_4$As$_3$ is well known as a one-dimensional Heisenberg antiferromagnet. \cite{Ochiai2, Kohgi}
In this compound, it has been discussed that although the CEF ground state has a large magnetic anisotropy, the system could be expressed as a one-dimensional isotropic $S$=1/2 Heisenberg antiferromagnet at a zero field. \cite{Shiba}
We also consider that in CeRu$_2$Al$_{10}$, a similar discussion may be applicable and a singlet ground state may be constructed. 
One of the characteristic features of CeRu$_2$Al$_{10}$ is that the magnitude of the Van Vleck term below $T_0$ shows a large anisotropy. 
In YbAl$_3$C$_3$, its difference is small.
Another mystery is the very high $T_0$ of CeRu$_2$Al$_{10}$.
The interaction between Ce ions should be an AFM one in order to form a singlet pair.
In order for the LRO with a singlet ground state to take place instead of the AFM order, the Peierls instability should be embedded in the crystal.
When $x$ increases in Ce$_x$La$_{1-x}$Ru$_2$Al$_{10}$, the system changes from the impurity Kondo to the Kondo semiconductor. 
How this variation with $x$ is associated with the LRO also remains to be explained. 
The small Curie term in $\chi$ below $T_0$ in La-doped samples should also be explained.
Many problems remain to be solved in this system in the future.

To conclude, we have investigated the thermal and transport properties of Ce$_x$La$_{1-x}$Ru$_2$Al$_{10}$ to clarify the origin of unknown phase transition at $T_0$=27 K in CeRu$_2$Al$_{10}$. 
We found that $T_0$ decreases with decreasing $x$ and that the $T_0$ of CeRu$_2$Al$_{10}$ is suppressed by a magnetic field along the $a$-axis.
These clearly indicate that the transition is ascribed to the interaction between Ce ions and that it has a magnetic nature. 
From the present results together with those in our previous papers, we propose that the singlet pair between Ce ions is formed below $T_0$. 
However, there still remains many problems to be explained. 
Especially, X-ray diffraction and inelastic neutron scattering should be performed to clarify the nature of the LRO.


\begin{thebibliography}{99}
\bibitem{Nishioka2} As for the determination of the crystal axes of this compound, we made a mistake in our previous study.\cite{Nishioka}  Correctly, the axis along which the largest magnetization and the smallest electrical resistivity are observed is the $a$-axis and that along which the intermediate magnetization and the largest electrical resistivity are observed is the $c$-axis. The authors thank Dr. Muro and Prof. Takabatake for pointing out our previous mistake on the determination of crystal axes.
\end{thebibliography}
\end{document}